\documentclass[a4paper,12pt]{article}
\usepackage{epsfig,amsmath,amsfonts}
\newcommand{\eqsection}{\makeatletter
   \@addtoreset{equation}{section}
   \renewcommand{\theequation}{\arabic{section}.\arabic{equation}}
   \makeatother}

\def\lal{&&\nqq {}}

\def\beq{\begin{equation}}
\def\eeq{\end{equation}}
\def\bear{\begin{eqnarray}}
\def\bearr{\begin{eqnarray} \lal}
\def\ear{\end{eqnarray}}
\def\earn{\nonumber \end{eqnarray}}

\def\d{\partial}

\def\const{{\rm const}}

\begin{document}

\begin{center}
\LARGE{\bf {An Emergent Universe with Dark Sector Fields in a
Chiral Cosmological Model}} 
\end{center}

\begin{center}

Beesham A., \\

{\it {Department of Mathematical Sciences, Zululand University \\
Private Bag X1001, Kwa-Dlangezwa 3886, South Africa }}

\bigskip
Chervon S.V.,\\%

{\it {Astrophysics and Cosmology Research Unit \\
School of Mathematical Sciences, University of KwaZulu-Natal \\
Private Bag X54 001 \\Durban 4000, South Africa}} and

\sl{\it {Ulyanovsk State Pedagogical University named after I.N.
Ulyanov, Ulyanovsk 432700, Russia}}

\bigskip
Maharaj S.D., \\

{\it {Astrophysics and Cosmology Research Unit \\
School of Mathematical Sciences, University of KwaZulu-Natal \\
Private Bag X54 001 \\Durban 4000, South Africa}}

\bigskip
Kubasov A.S., \\

\sl{\it {Ulyanovsk State Pedagogical University named after I.N.
Ulyanov, Ulyanovsk 432700, Russia}}

\end{center}

\small{ We consider the emergent universe scenario supported by a
chiral cosmological model with two interacting dark
sector fields: phantom and canonical. We investigate the general properties
of the evolution of the kinetic and potential energies as well as the development of the
equation of state with time. We present three models based on asymptotic
solutions and investigate the phantom part of the potential and chiral metric
components. The exact solution corresponding to a global emergent universe
scenario, starting from the infinite past and evolving to the infinite future,
has been obtained for the first time for a chiral cosmological model. The behavior
of the chiral metric components responsible for the kinetic interaction between the phantom
and canonical scalar fields has been analyzed as well.}

\vskip 0.2em

\small{\it Keywords: Cosmology, Emergent Universe, Dark Energy,
Phantom Field}

\section{Introduction}

Roughly a decade has passed since the appearance of the emergent universe (EmU)
scenario proposed by Ellis and Maartens \cite{ellmaa02}.
The discussions on the viability of this model include such topics as the
physical status of the emergent potential \cite{elmuts03}, existence and
stability of such a model \cite{mtle05}, confirmation of the existence of the
EmU solution in a Starobinsky model \cite{mukherjee05}, the influence
of exotic matter on the EmU evolution \cite{Mukherjee06}, and constraints on exotic matter needed for an EmU \cite{paul10}.
An extension of the EmU scenario for generalized Einstein gravity
containing an EmU in the brane world \cite{babach07,babach07-2} and chameleon,
$f(R)$ and $f(T)$ gravity theories \cite{chatto11} has been carried out. Ellis and Maartens \cite{ellmaa02}
had shown that the EmU is possible for closed Universe ($ \epsilon = +1$) if one considers a scalar field.
Debnath \cite{debnath08} showed that the EmU is possible for all values of $\epsilon = \{+1, -1, 0\} $ if one
 considers a phantom or tachyonic field. An EmU supported by a nonlinear sigma model with the potential of (self)interactions
(so-called {\it chiral cosmological model}) was proposed in the work
\cite{Beesham}.

It was shown in \cite{Beesham} that a nonlinear sigma model in its
most simple version with two components can support the
emergent universe scenario. From a logical point of view, it is very
suitable when we have only two scalar fields during the
time when the universe is undergoing slow expansion
from the minimal radius to the small radius. Actually one can consider the
self-interacting scalar field as a fluid of a special type with $
\rho = \textsc{K}+V,~ \textsc{K}:=\frac{1}{2}\dot{\phi}^2$ and
$p=\textsc{K}-V$. But it is definitely not possible to suggest the
presence of matter in the form of a fluid at this stage of the
universe's evolution: no particle, no fundamental forces.

Let us consider a single scalar field and a multiplet of them.
It is clear that we can always consider a scalar singlet as an
effective scalar field as the first stage of the phenomenon's
understanding. In the real situation, it is always a
composition of a few scalar fields: the relation between a multiplet
of scalar fields with geometrical (kinetic) interaction can be
described by an effective singlet with the relation \cite{ch97gc}:
$$
h_{AB}\varphi_{,\mu}^A\varphi_{,\nu}^B = \phi_{,\mu}\phi_{,\nu}
$$

Thus from the general relativity point of view, we have the same type of source
of the gravitational field, but more complicated dynamics in
the case of a multiplet of scalar fields. Therefore in the present
situation, we do not need to add any additional matter to the single
scalar field during the
universe emerging period ($ t\rightarrow -\infty $). We can state that a scalar field is
decomposed into at least two chiral fields and they are supporting
an EmU \cite{Beesham}.

Let us consider the second feature of a chiral cosmological model.
There are no ways generally speaking to present each scalar field
as an additive contribution in the energy density $\rho_{\sigma
}=\frac{1}{2}h_{AB}\varphi_{,\mu}^A\varphi_{,\nu}^B g^{\mu\nu}
+V(\varphi^C)$ because of the kinetic interaction. Indeed, e.g., in the case of a diagonal two-component chiral
cosmological model (in gaussian coordinates) we have
$$\rho_{\sigma }=\frac{1}{2}
\dot{\phi}^2+\frac{1}{2}h_{22}(\phi,\psi)\dot{\psi}^2+V(\phi,\psi)$$

Only in the case $h_{22}=h_{22}(\psi)$ and
$V(\phi,\psi)=V_1(\phi)+V_2(\psi)$ after redefinition of $\psi :
\dot{\tilde{\psi}}=\sqrt{h_{22}(\psi)}\dot\psi $ we arrive at the
multicomponent scalar field model with
\beq\nonumber
\rho_{\sigma}=\frac{1}{2} \dot{\phi}^2 + V_1(\phi) + \frac{1}{2}
\dot{\tilde{\psi}}^2+V_2(\tilde{\psi}) =
\rho_\phi+\rho_{\tilde{\psi}}
\eeq

In the present article we will show that the model with two scalar
(chiral) fields can describe an emergent universe in an asymptotic and exact
solutions basis. One of the fields (phantom one!) can be
considered as responsible for the evolution of the flat part of the
universe while another field is responsible for the appearance of the
curvature of the Universe.

\section{Chiral cosmological model}

We start from the action of a chiral cosmological model as the
action of a self-gravitating nonlinear sigma model (NSM) with the
potential of (self)-interaction $V(\varphi)$
\cite{ch95gc,ch95iv}:
\begin{equation}\label{1}
 S=\int\sqrt{-g}d^4x\left(\frac{R}{2\kappa}+
   \frac{1}{2}h_{AB}(\varphi)\varphi^A_{,\mu}\varphi^B_{,\nu}g^{\mu\nu}-V(\varphi)\right),
\end{equation}
where $g_{\mu\nu}(x)$~is the metric of the
space-time, $\varphi=(\varphi^1,\ldots,\varphi^N)$~is a
multiplett of the chiral fields (we use a notation
$\varphi^A_{,\mu}=\partial_{\mu}\varphi^A =\frac{\partial
 \varphi^A}{\partial x^\mu}$), and
$h_{AB}$~is the metric of the target space (chiral space) with
the line element
\beq\label{tsmet}
d s^2_{\sigma}=h_{AB}(\varphi)d\varphi^A d\varphi^B.
\eeq
The energy-momentum tensor for the model (\ref{1}) reads
\begin{equation}\label{2}
 T_{\mu\nu}=\varphi_{A,\mu}\varphi^A_{,\nu}-
 g_{\mu\nu}\left(\frac{1}{2}\varphi^A_{,\alpha}\varphi^B_{,\beta}g^{\alpha\beta}h_{AB}-
   V(\varphi)\right).
\end{equation}
The Einstein equations can be reduced to the form
\begin{equation}\label{3}
R_{\mu\nu}=\kappa(h_{AB}\varphi^A_{,\mu}\varphi^B_{,\nu}-g_{\mu\nu}V(\varphi))
\end{equation}

Varying the action (\ref{1}) with respect to $\varphi^C$, one can
derive the dynamic equation of a chiral field
\begin{equation}\label{4}
 \frac{1}{\sqrt{-g}}\partial_{\mu}(\sqrt{-g}\varphi^{,\mu}_A)-\frac{1}{2}\frac{\partial
   h_{BC}}{\partial\varphi^A}\varphi^C_{,\mu}\varphi^{B}_{,\nu}g^{\mu\nu}+V_{,A}=0,
\end{equation}
where $V_{,A}=\frac{\partial V}{\partial\varphi^A}$.
Considering the action (\ref{1}) in the framework of a
cosmological space, we arrive at a chiral cosmological model
\cite{ch02gc,ch00mg}.

Let us start from the case of a two component chiral cosmological
model as a source of the gravitational field with the target space
metric (\ref{tsmet}) in the form
\begin{equation}\label{5}
 d s^2_{\sigma}=h_{11}(\varphi,\psi)d\varphi^2+h_{22}(\varphi,\psi)d\psi^2
\end{equation}
Let us remember that we always make the choice
$h_{11}=\pm 1 $ which implies the choice of gaussian coordinates with
chosen signature. Henceforth we will consider $h_{11}$ only in the
above-mentioned sense, i.e., as a sign control symbol and not as a
function.

The energy-momentum tensor (\ref{2}) for the chiral metric
(\ref{5}) takes the form
\begin{equation}\label{6}
 T_{\mu\nu}=h_{11}\varphi_{,\mu}\varphi_{,\nu}+h_{22}\psi_{,\mu}\psi_{,\nu}-
 g_{\mu\nu}\left[\frac{1}{2}h_{11}\varphi_{,\rho}\varphi^{,\rho}+
 \frac{1}{2}h_{22}\psi_{,\rho}\psi^{,\rho}-V(\varphi,\psi)\right].
\end{equation}

The metric of a homogeneous and isotropic universe can be taken in the
Friedman--Robertson--Walker (FRW) form as
\begin{equation}\label{7}
 d s^2=d t^2-a(t)^2\left(\frac{d r^2}{1-\epsilon r^2}+r^2d\theta^2+r^2\sin^2\theta
   d\varphi^2\right).
\end{equation}

For the metric (\ref{7}),  the field equations of the two component
chiral cosmological model (\ref{4}) and Einstein's equations
(\ref{3}) can be represented in the form:
\begin{equation}\label{8}
h_{11}\ddot{\varphi}+3Hh_{11}\dot\varphi-\frac{1}{2}\frac{\partial
h_{22}}{\partial \varphi}\dot \psi^2+\frac{\partial
V}{\partial\varphi}=0,
\end{equation}
\begin{equation}\label{9}
3H(h_{22}\dot \psi)+\partial_t(h_{22}\dot
\psi)-\frac{1}{2}\frac{\partial h_{22}}{\partial \psi}\dot
\psi^2+\frac{\partial V}{\partial\psi}=0,
\end{equation}
\begin{equation}\label{10}
H^2=\frac{\kappa}{3}\left[\frac{1}{2}h_{11}\dot\varphi^2+\frac{1}{2}h_{22}\dot\psi^2+V\right]-\frac{\epsilon}{a^2},
\end{equation}
\begin{equation}\label{11}
\dot
H=-\kappa\left[\frac{1}{2}h_{11}\dot\varphi^2+\frac{1}{2}h_{22}\dot\psi^2\right]+\frac{\epsilon}{a^2}.
\end{equation}

This system of equations is the system of differential equations
of second order with three unknown variables: two chiral fields
$\varphi$ and $\psi$, and potential $V$. In accordance with the method
of fine tuning of the potential \cite{ch97mono}, the law of
evolution of the Universe  $a=a(t)$ is specified. The metric of a
target space is not fixed as it is traditionally accepted, giving us the
freedom of adaptation to resolving the problem. Making a simple
algebraic conversion of the Einstein equations (\ref{10})--(\ref{11}),
we find their useful implication:
\begin{equation}\label{12}
 \frac{1}{2}h_{11}\dot\varphi^2(t)+\frac{1}{2}h_{22}(t)\dot\psi^2(t)=
 \frac{1}{\kappa}\left[\frac{\epsilon}{a^2}-\dot
   H\right],
\end{equation}
\begin{equation}\label{13}
 V(t)=\frac{3}{\kappa}\left(H^2+\frac{1}{3}\dot
   H+\frac{2}{3}\frac{\epsilon}{a^2}\right).
\end{equation}

To obtain the exact solutions presented in the article, we demand
that mappings $\psi(t), \varphi(t)$ and $t(\psi), t(\varphi)$  are
single valued and simple (not transcendental).

It is convenient to search for solutions with the following form
of metric components $h_{22}$ and total potential $V_{tot}$
\begin{equation}\label{15}
h_{22}(\varphi,\psi)\equiv h_{22}(\varphi),\quad
V_{tot}(\varphi,\psi)=V_1(\varphi)+e^{f(\varphi)}V_2(\psi)+V_3(\psi).
\end{equation}
respectively.

Solving the  equation (\ref{9}) with the substitution (\ref{15}) we
have a problem. In this case equation (\ref{9}) reads
\begin{equation}\label{110}
\sqrt{h_{22}}\frac{\d V_2}{\d \psi}+\frac{\d V_3}{\d
\psi}=-3Hh_{22}\dot\psi+\d_t(h_{22}\dot\psi).
\end{equation}
To avoid the problem of solving this equation, we shall consider
the case when $h_{22}\rightarrow 0$ and $V_3\rightarrow 0$. In
this situation equation (\ref{110}) tends to zero.

\section{Emergent Universe scenario with two dark sector fields}

A nonlinear sigma model
has already been considered as the source of the emergent universe 
\cite{Beesham}. Here we consider the 2-component NSM and we
choose the scale factor in the most general form
\cite{Mukherjee06}
\begin{equation}\label{a_emu}
a(t)=A(\beta+e^{\alpha t})^m,~~\alpha >0,~~\beta >0,~~m>0.
\end{equation}

Let us carry out an analysis of the general evolution of the EmU and
physical interpretation of the model's parameters. The EmU started off
from the radius $a_i>M_P^{-1}$ \cite{ellmaa02} in the infinite
past $t \rightarrow -\infty $. Using this asymptote, we find
that $A=\frac{a_i}{\beta^m}$. Starting from the radius $a_i$, the
scale factor of the Universe then increases until the epoch
$t=0$. Let us denote this radius as $a_s$. When we consider the
evolution from $t \rightarrow -\infty $ to $t \rightarrow +\infty
$ and there is no singularity for the $t=0$ epoch, we can choose
the time that inflation starts. For example, we can choose
$\dot{a}=0.001 $ as the moment when the velocity of expansion of the Universe
is about 1/1000-th part of the velocity of light. The number
of e-folds to the $t=0$ moment can be estimated as
\beq\label{e-f-0}
\textsc{N}_s=\ln \frac{a_s}{a_i}=m\ln \left(1
+\frac{1}{\beta}\right)
\eeq
From (\ref{e-f-0}) we conclude that the parameter $m$
reduces of number of e-folds until the zero moment. Also we can express the
parameter $\beta $ in terms of the e-fold number $\textsc{N}_s $ as:
$$
\beta =\frac{1}{\exp \left({\textsc{N}_s}/{m}\right)-1}.
$$
Obviously, until the end of inflation, if we will have the radius of
the Universe being $a_e$, then the number of e-folds from inflation will be
$$
\textsc{N}_{inf}=\ln\frac{a_e}{a_s}.
$$
The total number of e-folds will thus be the sum of $\textsc{N}_s$ and
$\textsc{N}_{inf}$. Note that for the EmU scenario, we have eternal
inflation and exit from it may be considered in the standard
inflationary scenario way: decay of scalar fields and creation of
particles, reheating, etc \cite{ellmaa02}.

\subsection{Evolution of the total kinetic energy and the potential}

Now let us turn to an analysis of the implications of Einstein's
equations (\ref{12}), (\ref{13}) which for the EmU scenario take the
form:
\begin{equation}\label{66}
 \frac{1}{2}h_{11}\dot\varphi^{2}(t)+\frac{1}{2}h_{22}\dot
 \psi^{2}(t)=\frac{1}{\kappa}\left(-\frac{m\alpha^2\beta e^{\alpha
       t}}{(\beta+e^{\alpha t})^2}+\frac{\epsilon}{A^2(\beta+e^{\alpha
       t})^{2m}}\right),
\end{equation}
\begin{equation}\label{67}
V(t)=\frac{1}{\kappa}\left(\frac{m\alpha^2 e^{\alpha
t}(3me^{\alpha t}+\beta)}{(\beta + e^{\alpha
t})^2}+\frac{2\epsilon}{A^2(\beta+e^{\alpha t})^{2m}}\right).
\end{equation}

From these equations, we can find the general behavior for the evolution of
the total kinetic energy $\textsc{K} $ and total potential $V_{tot}$.
To this end, let us consider asymptotes $t \rightarrow \pm \infty$
and zero time value (with $\kappa
=1,~~\epsilon=1$) for the kinetic energy:
\bear
\textsc{K}_{(t=-\infty)} = \frac{1}{a_i^2}\\
\textsc{K}_{(t=0)}= - \frac{m\alpha^2\beta}{(\beta+1)^2}+\frac{1}{a_i^2 \exp \left(\textsc{N}_s/2\right)}\\
\textsc{K}_{(t= +\infty)}= 0
\ear
Let us mention that during evolution, the  kinetic energy may be less
then zero, because of the phantom character of one of the chiral
fields. The graphs for special values of the parameters are displayed in Fig.1.

\begin{figure}[h!]
\center
\includegraphics[width=3.25 in]{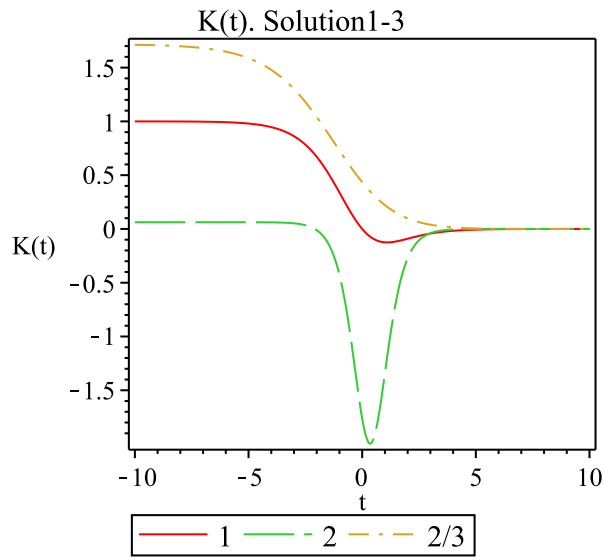}
\caption{The total kinetic energy evolution} \label{ris:1}
\end{figure}

\begin{figure}[h!]
\center
\includegraphics[width=3.25 in]{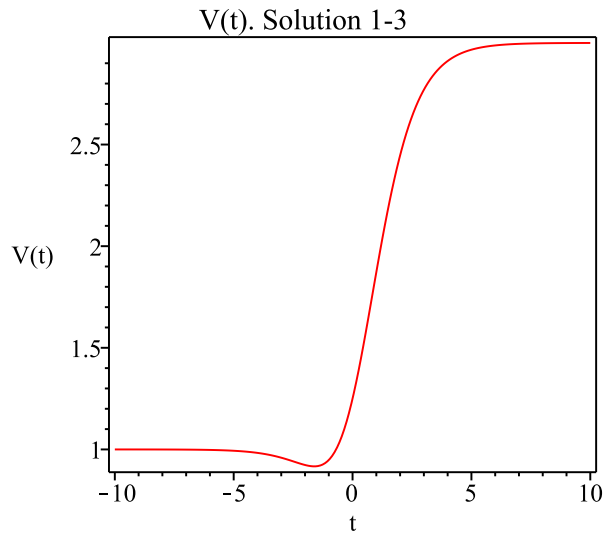}
\caption{Evolution of the potential} \label{ris:2}
\end{figure}

For the total potential we have at the asymptotes $t \rightarrow \pm \infty$
and zero time value:

\bear
{V_{tot}}_{(t=-\infty)}= 2\textsc{K}_{(t=-\infty)} = \frac{2}{a_i^2}\\
{V_{tot}}_{(t=0)}=\left(\frac{m\alpha^2(3m+\beta)}{(\beta + 1)^2}+\frac{2\beta^{2m}}{a_i^2(\beta+1)^{2m}}\right) \\
{V_{tot}}_{(t=+\infty)}=3m^2\alpha^2
\ear

For the special choice of parameters, we have a local minimum
for negative time (Fig.2).

\subsection{Evolution of the equation of state}

Let us consider the general evolution of the EOS for the EmU scenario.
With the definition of the scale factor (\ref{a_emu}), one can find
$$
w_{tot}=\frac{p}{\rho}=\frac{\textsc{K}-
V_{tot}}{\textsc{K}+V_{tot}}
$$
Let us consider the asymptotes. At times when $t \rightarrow
-\infty $ the EOS has the limiting form:
\beq
w_{tot}(-\infty)=-\frac{1}{3}
\eeq

When $t \rightarrow 0 $ one can easily obtain a general formula
for the EOS. We will consider the result with the assumptions that
$m=1$ and $\beta =1 $. In this case we have:
\beq
w_{tot}(0)=-\frac{5\alpha^2a_i^2+1}{3\alpha^2a_i^2+3}.
\eeq
This give us an EOS close to $-1$, if $\alpha^2a_i^2$ is of the order of unity.

The case $t \rightarrow +\infty $ gives us the direct answer that
$$
w_{tot}(+\infty)=-1.
$$
This means that at late times the Universe turns to the $\Lambda
$ dark energy scenario.

This analysis confirms that it is the dark energy fields that give rise
to the EmU scenario.

\subsection{The number of e-foldings}

In our work we have obtained exact solutions which gives us
the possibility to use exact inflation techniques \cite{chefom08}
instead of the slow-roll approximation.

Let us follow the prescription for the superpotential
construction in the case of a canonical single scalar field
\cite{yurov11}. If we choose the superpotential as the total
energy potential in the form
\beq
\textsc{W}=V+\textsc{K}= V +
\frac{1}{2}h_{11}\dot\varphi^2(t)+\frac{1}{2}h_{22}(t)\dot\psi^2(t)
\eeq
then the linear combination of the equations for the chiral fields leads to
the following relation
\beq
3H\left(
h_{11}\dot\varphi^2(t)+h_{22}(t)\dot\psi^2(t)\right)=-\frac{d\textsc{W}}{dt}
\eeq
This relation allows us to calculate the e-folding
number for each exact solution with the formula
\beq\label{N-phy-psi}
\textsc{N}= -\int_{t_i}^{t_f}
\frac{\textsc{W}-\frac{\epsilon}{a^2}}{\frac{d\textsc{W}}{dt}}\left(2\textsc{K}\right)dt
\eeq
As we have two fields depending on cosmic time $t$, it will be easier to
integrate over $t$.
Numerical integration in (\ref{N-phy-psi}) may also be
applied.

Let us mention here that for the power spectrum and spectral
indexes analysis, the exact inflation approach  \cite{chefom08} can
be applied for the model under consideration as well.

Now we will present three models based on asymptotic solutions
when $t \rightarrow \pm \infty$  and $V_3(\psi)$  tends to zero (\ref{15}),
and one exact solution describing the global evolution of the EmU.
These solutions have been obtained with the use of the freedom
in choice of the  chiral metric component $h_{22}$, viz., we consider
the case when the first chiral field $\varphi $ corresponds to the
second term in eq. (\ref{12}), i.e., we set
\beq\label{phidot}
\frac{1}{2}h_{11}\dot\varphi^2(t)= -\frac{1}{\kappa}\dot{H}
\eeq
From this relation it is easy to see that since $ \dot {H}>0
$ for the EmU scenario, we need $h_{11}<0$. Without loss of generality
we can choose gaussian coordinates with $h_{11}=-1.$
This means that the first chiral field should be a phantom one. We connect the
second chiral field  with the curvature of the Universe, viz.,
we set
\beq\label{psidot}
\frac{1}{2}h_{22}(t)\dot\psi^2(t)=
\frac{1}{\kappa}\left[\frac{\epsilon}{a^2}\right].
\eeq
In the case
of the EmU, we obtain $h_{22}>0$, i.e., the second chiral field $\psi
$ should be considered as an ordinary (canonical) field.

\section{The Model\, 1}

For simplicity, let us choose the Newton gravitational
constant $\kappa =1 $.
Thus by integrating equation (\ref{phidot}), we obtain for the first
(phantom!) field the solution:

\begin{equation}\label{varphi-1}
2\sqrt{2m}\tilde{\varphi}:=(\varphi(t)-\varphi_i)=
2\sqrt{-2mh_{11}}
\arctan\left(\frac{e^{\frac{\alpha}{2}t}}{\sqrt{\beta}}\right)=
2\sqrt{2m}
\arctan\left(\frac{e^{\frac{\alpha}{2}t}}{\sqrt{\beta}}\right),
\end{equation}
where $\varphi_i $ stands for the value of the scalar field at the time
$-\infty $ . We will restrict our consideration to the period
$-\frac{\pi}{2} <\tilde{\varphi}<\frac{\pi}{2} $ which allows for the time to run from  $-\infty$ to $+\infty $. Let us remember that we can
represent the parameter
$A$ as the combination of $a_i$ (initial size of the EmU), and
parameters $\beta $ and $m$ from the following relation:
$A=\frac{a_i}{\beta^m}$. Let us note also that the solution (\ref{varphi-1})
is valid for all four models.

The evolution of the second canonical field can be considered as given.
Then we can calculate the chiral metric component $h_{22}$
from (\ref{psidot}). For the first model we choose the dependence of
$\psi $ on cosmic time as follows

\begin{equation}\label{psi-1}
\tilde\psi =\psi(t)-\psi_i=\frac{\sqrt{2}}{\zeta}e^{\zeta t},~~\zeta=\const,
\end{equation}
Then the chiral metric component $h_{22}$ can be expressed in the form:
\begin{equation}
h_{22}(\tilde{\varphi})= \frac{1}{a_i^2}
\left[\beta\tan^2\left(\frac{\varphi(t)-\varphi_i}{
2\sqrt{2m}}\right)\right]^{-\frac{2\zeta}{\alpha}}\cos^{4m}\left(\frac{\varphi(t)-\varphi_i}{
2\sqrt{2m}}\right) =\frac{1}{a_i^2}
\left(\beta\tan^2(\tilde{\varphi})\right)^{-\frac{2\zeta}{\alpha}}\cos^{4m}(\tilde{\varphi}),
\end{equation}
The $V_1$ part of the potential takes the form:
\begin{equation}\label{w-1}
V_1(\varphi)=\frac{m\alpha^2}{4}\sin^2(2\tilde{\varphi}
)[3m\tan^2(\tilde{\varphi})+1].
\end{equation}

It is worth stressing here that the solution for $V_1 $ (\ref{w-1}),
as well as for $\tilde{\varphi} $ (\ref{varphi-1}), are valid for all
models below. The dependence of the shape of the potential $V_1$ vs $\tilde{\varphi}$
related to the parameter $m$ value is displayed in Fig. 3
and Fig. 4.

We will consider the $V_2$ part of the potential in the limiting case when
$ t \rightarrow \pm \infty $. Thus $V_2$ and its multiplier are:

\begin{equation}\label{f-vp}
f(\varphi)=\frac{1}{2}\ln\left(h_{22}(\varphi)\right),\quad
\end{equation}
\begin{equation}
 V_2(\psi)=\frac{2\beta^m}{a_i}\frac{\left(\frac{\zeta}{\sqrt{2}}\tilde\psi\right)}{\left[\beta+
 \left(\frac{\zeta}{\sqrt{2}}\tilde\psi\right)^{\frac{\alpha}{\zeta}}\right]^m},
\end{equation}

Let us investigate the case of positive and negative $\zeta $.
For the sake of simplicity let $ \zeta =1$.

The general features of $h_{22}$ are as follow (see Fig.5). We consider
$h_{22}$ as an even function to prevent a crossing of the phantom
boundary (this case may be studied separately). During times when
$ t \rightarrow \pm\infty $, the chiral metric component $h_{22}
\rightarrow 0$. This can be interpreted as the absence of
influence from the second chiral field at the mentioned stages (as one
can see in spite of the growth of the second field itself).
Nevertheless during times close to zero, the value of $h_{22}$
tends to infinity. This means that the second field $\psi $ plays
an important role during the inflationary period and acts as the inflaton.

The second possibility is $ \zeta =-1$.

The situation in this case is different (see Fig. 6). During the inflationary
stage, when time tends to zero, $h_{22}$ tends to zero and almost
cancels the influence of the second field $\psi $ during times
close to zero. This means that the first chiral field $\varphi $
acts as the inflaton. During the period when time tends to $\pm\infty
$, the chiral metric component $h_{22}$ takes maximum values. This
means that the second field $\psi $ plays an important role at
infinite times.

There is one more possibility to have a solution with two maximums and
one minimum by matching parameters of the model. Then $h_{22}$
tends to zero, when $t \rightarrow \pm\infty $ and $t \rightarrow
\pm 0 $. Thus the effect of the second field $\psi $ is cancelled
during those times.

Let us turn our attention to the potential $V_1(\varphi )$.
Its behavior looks like the case of $h_{22}$ with
$\zeta =-1$ in Fig.6. The difference is in the amplitude.

\begin{figure}[h!]
\center
\includegraphics[width=3.25 in]{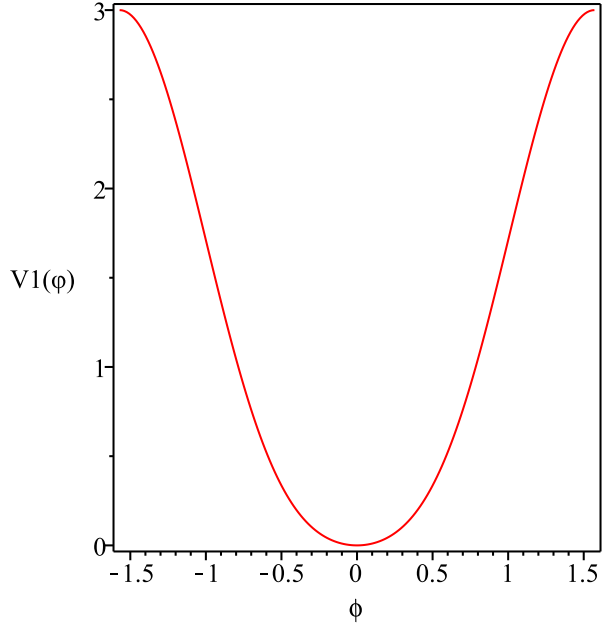}
\caption{The potential vs the field $\varphi$, with parameters:
m=1, $\alpha=1$} \label{ris:3}
\end{figure}

\begin{figure}[h!]
\center
\includegraphics[width=3.25 in]{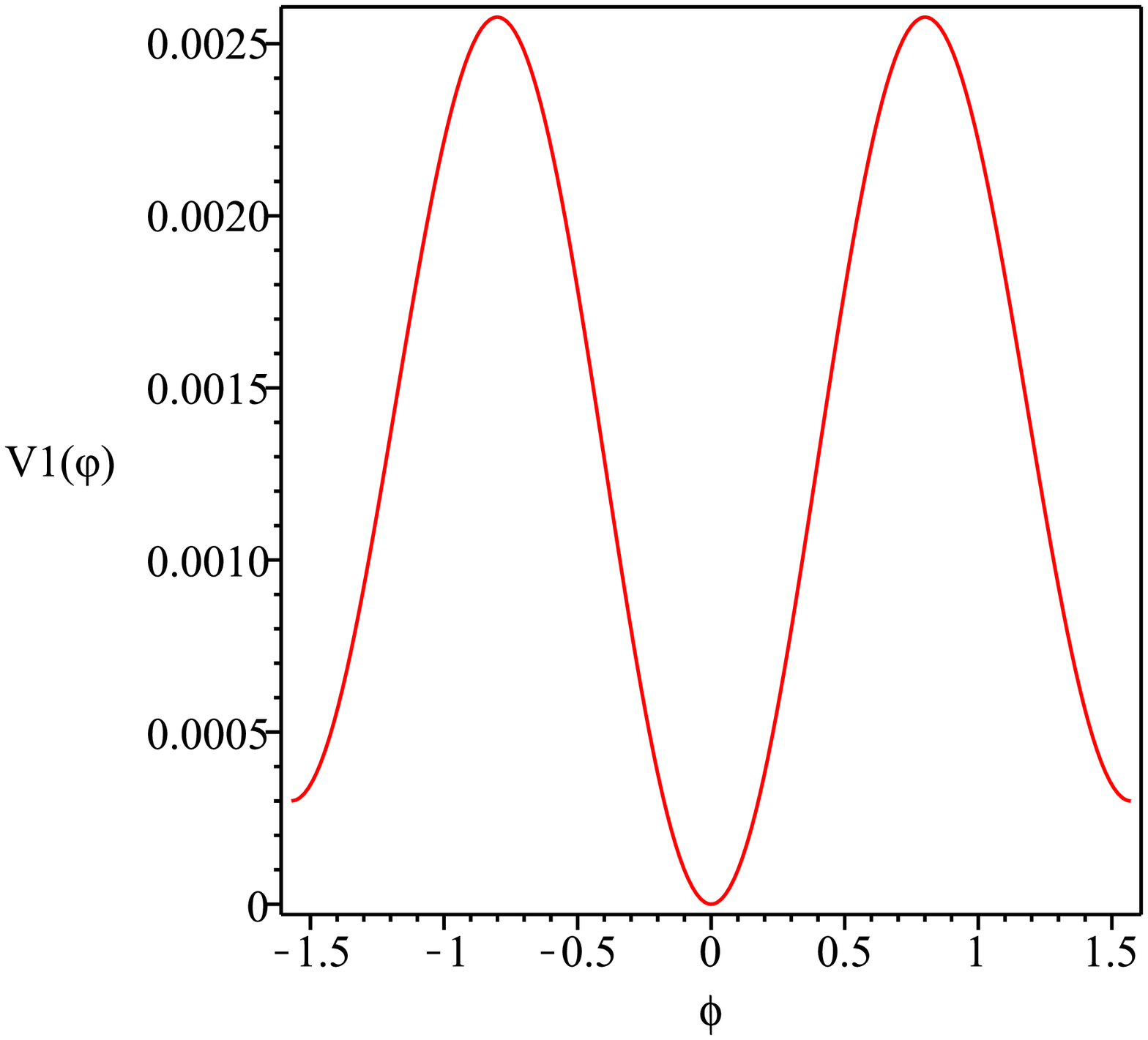}
\caption{The potential vs the field $\varphi$, with parameters:
m=0.01, $\alpha=1$} \label{ris:4}
\end{figure}

\begin{figure}[h!]
\center
\includegraphics[width=3.25 in]{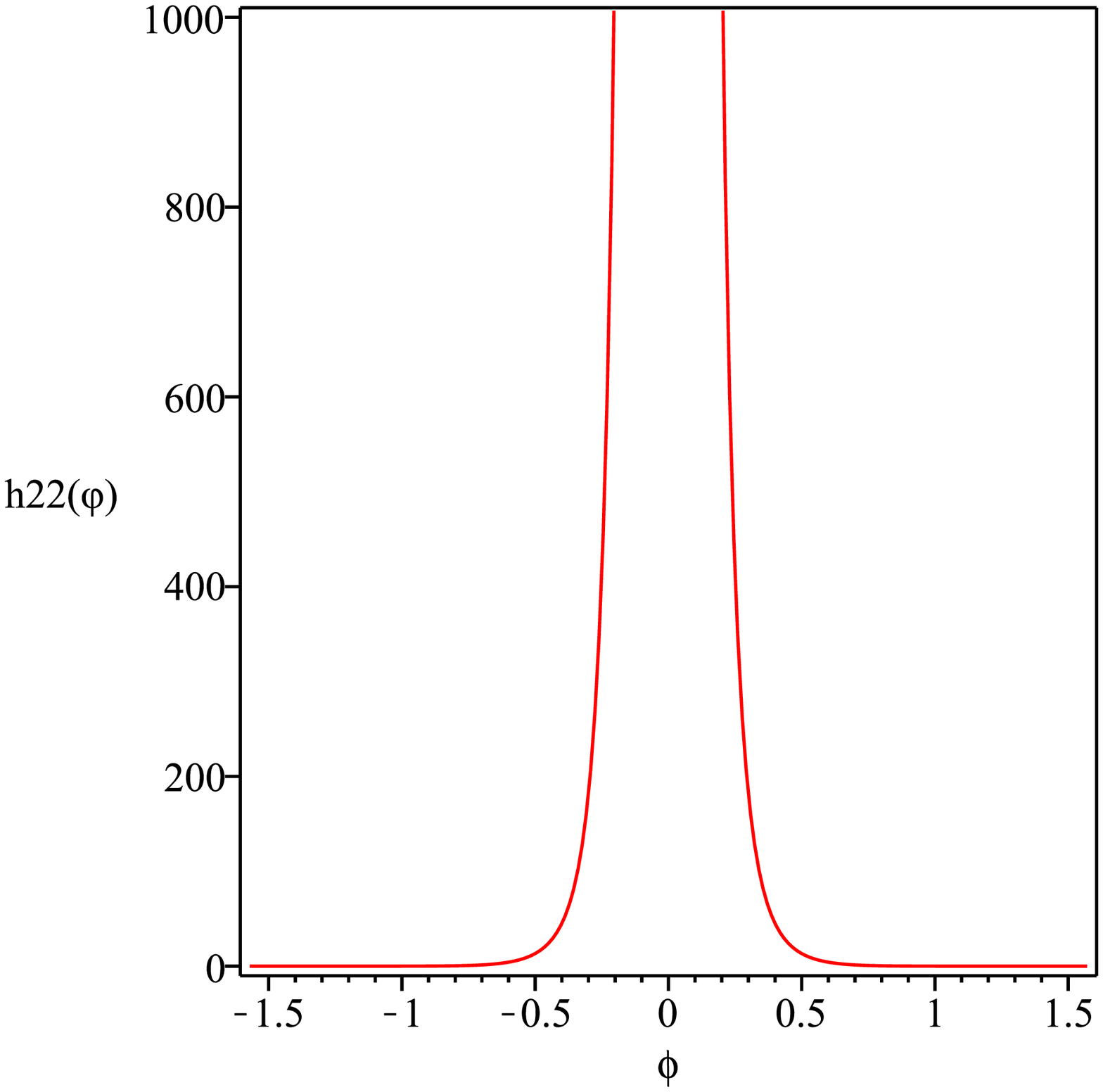}
\caption{Model 1. The $h_{22}(\varphi)$ component vs the field
$\varphi$ with parameters: A=1, m=1, $\alpha=1$, $\beta=1$,
$\zeta=1$} \label{ris:5}
\end{figure}

\begin{figure}[h!]
\center
\includegraphics[width=3.25 in]{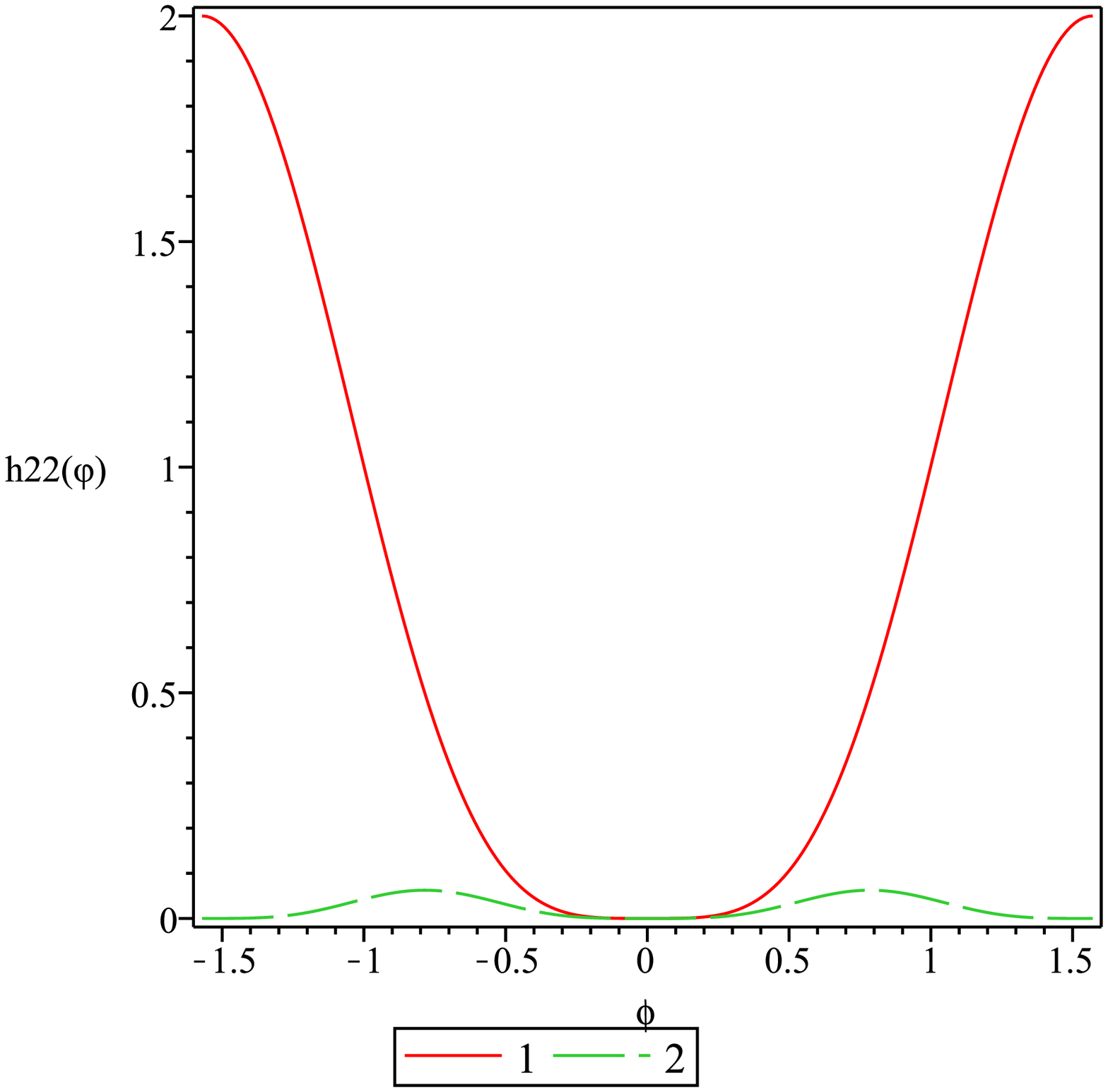}
\caption{Model 1. The $h_{22}(\varphi)$ component vs the field
$\varphi$ with $\zeta=-1$ and
parameters changing} \label{ris:6}
\end{figure}

\section{The Model\, 2}
The model 1 obeys the property that the value for the second chiral field
$\psi $ tends to infinity in the infinite future.
From the point of view of background dark sector fields
\cite{chepan2010}, it is preferable that the values of the fields
are restricted. Therefore we are looking for the solution of a kink type
for chiral fields. Model 2 gives us such a solution. We
will keep the solution for the chiral field $\tilde{\varphi}$ in the
same form as in eq. (\ref{varphi-1}) as well as the solution
for the $V_1$ part of the potential (\ref{w-1}). This is possible
because of our choice to keep for the first chiral field the
responsibility for the spatially-flat part of the Universe. Let us
mention once more that in the EmU, the first chiral field should be a phantom one
because of relation $ h_{11}\dot\varphi^2=-\dot{H}$ and the fact that
$\dot{H}>0$ for the scale factor (\ref{a_emu}). We can also state
that the $V_1$ part of the potential is determined by the spatially-flat
part of the Universe. When the multiplier $ e^{f(\varphi)}$ is fixed,
we have a dependence of the $V_2$ part of the potential on the
second field $\psi $.

The solutions obtained for {\bf model 2} are
\begin{equation}
K_1 \tilde{\psi}:=\psi(t)-\psi_i=K_1\arctan\left(\lambda t\right)
\end{equation}
\beq
h_{22}=\frac{2}{a_i^2 K_1^2 \lambda^2}\cos^{4m}(\tilde{\varphi})
\left[1+\frac{\lambda^2}{\alpha^2}\ln^2 (\beta \tan^2
\tilde{\varphi})  \right]^2
\eeq
where $K_1$ is an arbitrary constant.
The potential $V_2 (\psi)$ once again in the limiting cases
$t \rightarrow \pm \infty $, after some calculations, takes the form
\beq
V_2 =\frac{\sqrt{2}\beta^{m} \lambda
K_1}{a_i}\cos^{2}(\tilde{\psi})
\left[\beta+\exp\left(\frac{\alpha}{\lambda}\tan
\tilde{\psi}\right) \right]^{-m}
\eeq

We can see from Fig. 7 and Fig. 8 the possible types of the $h_{22}$ evolution.
The first possibility in Fig. 7 gives us a good example
of a singularity-free solution. The influence of the second field at
infinite times (past and future infinity) tends to zero, i.e., the dynamics
of the Universe is determined by the first (phantom) field $\varphi $.
During the inflationary period, the role of the second field $\psi $
became essential. The shape of the chiral metric component $h_{22} $
displayed in Fig. 8 provides us with some other information
about the second field, viz., the field $\psi $ plays an important
role at infinite times and during the inflationary period where $h_{22} $ tends
to infinity.

\begin{figure}[h!]
\center
\includegraphics[width=3.25 in]{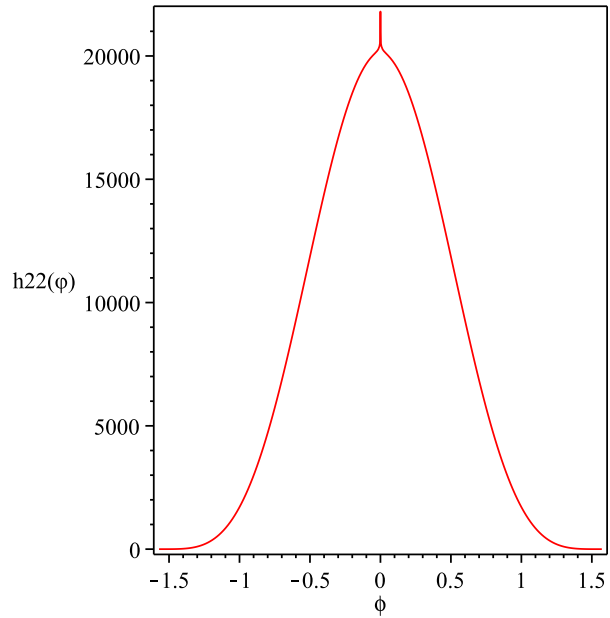}
\caption{Model 2. The $h_{22}(\varphi)$ component vs the field
$\varphi$ with parameters: A=1, K=1, m=1, $\alpha=1$, $\beta=1$,
$\lambda=0.01$} \label{ris:7}
\end{figure}

\begin{figure}[h!]
\center
\includegraphics[width=3.25 in]{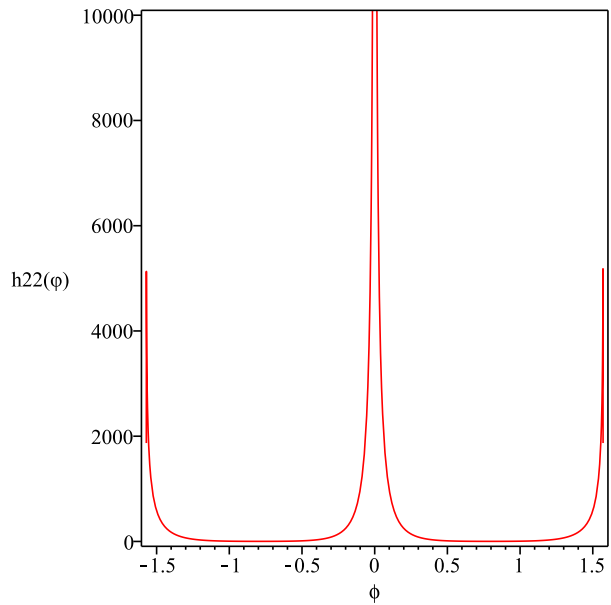}
\caption{Model 2. The $h_{22}(\varphi)$ component vs the field
$\varphi$ with parameters: A=1, K=1, m=0.1,$\alpha=1$, $\beta=1$,
$\lambda=1$} \label{ris:8}
\end{figure}

\section{The Model\, 3}

As we mentioned earlier, {\bf model 1} allows infinite values for the
second chiral field $\psi $. From the position of background dark sector
fields
\cite{chepan2010}, it would be more suitable if the values of the fields
could be restricted. Therefore we seek solutions of the kink type
for chiral fields. {\bf Model 2} gives us such a solution. Let us
consider the solution for the chiral field $\tilde{\varphi}$ in the
same form as in eq. (\ref{varphi-1})
\begin{equation}
2\sqrt{2m}\tilde{\varphi}:=\varphi(t)-\varphi_i= 2\sqrt{2m}
\arctan\left(\frac{e^{\frac{\alpha}{2}t}}{\sqrt{\beta}}\right),
\end{equation}
The same result holds for the $V_1$ part of the potential, viz.,
\begin{equation}
V_1(\varphi)=\frac{m\alpha^2}{4}\sin^2(2\tilde{\varphi}
)[3m\tan^2(\tilde{\varphi})+1],
\end{equation}
As a possible evolution of the second (canonical) chiral field we
can consider a few possibilities, firstly, connected with the
solution investigated in {\bf model 1}. To avoid an infinite
increase of the solution (\ref{psi-1}) when $t \rightarrow \pm \infty,
$ let us choose
\begin{equation}\label{44}
K_1 \tilde{\psi}:=\psi(t)-\psi_i=K_1\arctan\left(\frac{\sqrt{2}}{\zeta} e^{\zeta t}\right)
\end{equation}
where $K_1 $ is arbitrary constant.

With the choice (\ref{44}), the chiral metric component $h_{22}$ can be
expressed in the form:
\begin{equation}
h_{22}(\tilde{\varphi})=\frac{1}{a_i^2
K_1^2}\left[1+\frac{2}{\zeta^2}
\left(\beta\tan^2\tilde{\varphi}\right)^{\frac{2\zeta}{\alpha}}
\right]^2\left[\beta\tan^2\tilde{\varphi}\right]^{-\frac{2\zeta}{\alpha}}\cos^{4m}\tilde{\varphi},
\end{equation}
After some calculations, we can find the $V_2$ part of the potential as
\beq
V_2=\frac{K_1\zeta\beta^m}{\sqrt{2}a_i}\sin 2
\tilde{\psi}\left[\beta+\left( \frac{\zeta}{\sqrt{2}}\tan
\tilde{\psi}\right)^{\frac{\alpha}{\zeta}}\right]^{-m}
\eeq
The function $f(\varphi)$ is determined by the same relation
(\ref{f-vp}):$
f(\varphi)=\frac{1}{2}\ln\left(h_{22}(\varphi)\right).$

The component $h_{22}$ may have a very different shape depending
on the parameters of the model. We can find shapes similar to that
for {\bf model 2}. But here we can find the chiral metric component
$h_{22}$ in principally new forms, displayed in figures 9-10
which show the possible dependence on $\varphi $.


\begin{figure}[h!]
\center
\includegraphics[width=3.25 in]{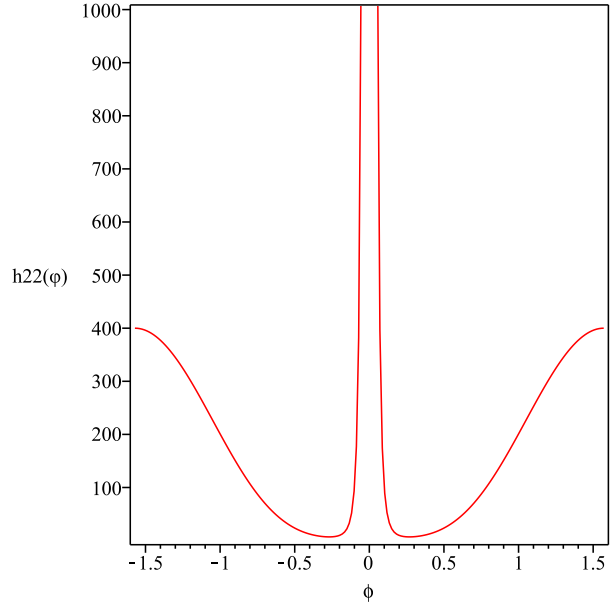}
\caption{Model 3. The $h_{22}(\varphi)$ component vs the field
$\varphi$ with parameters: A=1, K=1, m=1,$\alpha=1$, $\beta=10$,
$\zeta=1$} \label{ris:9}
\end{figure}

\begin{figure}[h!]
\center
\includegraphics[width=3.25 in]{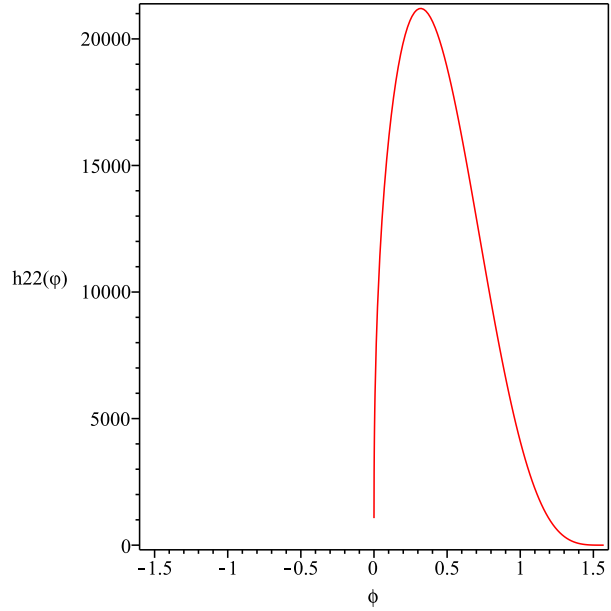}
\caption{Model 3. The $h_{22}(\varphi)$ component vs the field
$\varphi$ with parameters: A=1, K=1, m=1,$\alpha=1$, $\beta=1$,
$\zeta=0.1$} \label{ris:10}
\end{figure}

\section{The Model\, 4}

We have considered solutions which are valid for times $t \rightarrow \pm \infty $.
To connect the results with the entire evolution of the universe, we need an exact solution for the model.
An example of such a solution can be found with the assumption that the second canonical field
$\tilde\psi $ is proportional to the cosmic time $t$, viz., if we choose $\tilde\psi=\sqrt{2}t $, then
we have the
same solution for the first phantom field $\tilde\varphi $ (\ref{varphi-1}) and the $V_1$ part of the potential
(\ref{w-1}). The $V_3$ part of the potential vanishes. Thus the exact solution which is valid for
the entire evolution of the Universe reads

\begin{equation}
2\sqrt{2m}\tilde{\varphi}:=\varphi(t)-\varphi_i= 2\sqrt{2m}
\arctan\left(\frac{e^{\frac{\alpha}{2}t}}{\sqrt{\beta}}\right),
\end{equation}

\begin{equation}
V_1(\varphi)=\frac{m\alpha^2}{4}\sin^2(2\tilde{\varphi}
)[3m\tan^2(\tilde{\varphi})+1],
\end{equation}

\begin{equation}
\tilde\psi=\sqrt{2}t
\end{equation}

\begin{equation}
h_{22}(\varphi)=\frac{[\cos^2(\tilde\varphi)]^{2m}}{a_i^2}
\end{equation}

\begin{equation}
V_2(\psi)=\frac{2}{A(\beta
+e^{\frac{\alpha\tilde\psi}{\sqrt{2}}})^{m}},~~~V_3(\psi)=0
\end{equation}

Let us compare the asymptotes of exact model 4 with the models 1-3.
When $t \rightarrow +\infty $ we can state that for the exact model
$h_{22} $ and $V_2$ tend to zero. We can find the same asymptote  under
special choices of parameters for the model 1 (see Fig. 5 and Fig. 6),
for the model 2 (see Fig.7) and for the model 3 (see Fig. 10). As for the
negative time asymptote $t \rightarrow -\infty $, we have $h_{22} \rightarrow 0 $
which agrees with  models 1-3, but for $V_2$ we have $V_2 \rightarrow \frac{2}{a_i}$.
The last asymptote does not exist for  models 1-3. The potential part $V_2$ can tend
to zero or to infinity.

\section{Discussions}

In this article, the ideas presented in the work
\cite{Beesham} have been further developed. In
\cite{Beesham}, we considered the simplified equations for
the two component nonlinear sigma model during the time when $t
\rightarrow -\infty $ and also the simplified metric of the EmU during the
inflationary period. Here we have presented models 1-3 with asymptotic
solutions for positive and negative infinite times and model 4 which
contains the exact
solution with two chiral fields as some kind of dark sector
fields. Model 4 describes the EmU from $t\rightarrow-\infty $
until the end of inflation, and  coincides with the asymptotes of models 1-3
for those times.

If we compare the solutions obtained in the present article with
those presented in \cite{Beesham}, it becomes clear that
asymptotically they may coincide under a suitable choice of the initial
values for the first chiral field. But for the chiral metric component
$h_{22}$ and for the second chiral field $\psi $ the results are very
different. Component $h_{22}$ tends to a constant value when $t \rightarrow -\infty $,
while $\psi \rightarrow -\infty $ as well. This is not very clear from a
physical point of view: in the infinite past there exists
a scalar field with the amplitude tending to minus infinity.

We can see another feature from the explicit solution presented in \cite{elmuts03} for a scalar field singlet. We mention here
that the exact solutions for the EmU obtained by the authors in \cite{elmuts03}
with the help of the fine turning method of the potential was suggested
for the fist time in the work \cite{ellmad91} and then developed
in the works \cite{chzhsh97pl}, \cite{chzhsh98zetf}. Afterwards, this method
was reopened in the work \cite{paddy02} and applied to tachyonic matter.

The solution in \cite{elmuts03} consists of two parts: asymptotically
a static Einstein Universe when $t \rightarrow -\infty $ and at late times
the scale factor has exponential behavior
$a(t) \propto \exp (\alpha t)$.

If we consider the solutions for the potential and the scalar field,
we find an unpleasant situation with the scalar field: it tends
to $-\infty $, and  the potential tends to a constant value.
We have the same situation with the $V_1$ part of the potential for the
models 1-3 presented in this article. For the {\bf model 1}, we have
constant values for $t \rightarrow -\infty $ and an infinite increase of the
second field for $t \rightarrow \infty $. But for {\bf model 2} and
{\bf model 3} this unpleasant behavior was corrected. Thus we suggested
models which are reasonable from a physical point of view:
the chiral fields (the first is phantom one, while the second is canonical)
are evaluated during the global evolution of the Universe and they are restricted
in amplitudes. The kinetic interaction between these two fields has
the feature that during the inflationary stage they may have infinite values.
This fact may be interpreted as a dominant influence of the second canonical
scalar field during the inflationary stage. Nevertheless, for a special choice
of parameters, the amplitude of the second field may have a large but finite
value.

Let us stress here that we have obtained for the first time an exact solution for
the EmU supported by two chiral fields in {\bf model 4}.
When $t \rightarrow -\infty $ we have $\psi \rightarrow -\infty $, the chiral metric
component $h_{22} \rightarrow 0$ and the $V_2$ part of the potential tends to
the constant: $V_2 \rightarrow \frac{2}{a_i} $. Nevertheless, if we take
into account the multiplier $e^{f(\varphi)}=\sqrt{h_{22}}$ of the $V_2$  which
tends to zero, effectively the situation will be as in the case of
models 1-3. When $t \rightarrow +\infty $, we have $\psi \rightarrow +\infty $ and  the chiral metric
component $h_{22}$  and the $V_2$ part of the potential tend to zero which is
in agreement with models 1-3.

Summing up the result of the article we can state that the exact solutions of
the model 4 we found here can support the EmU during the entire evolution of
the Universe,
starting from the infinite past and evolving to the end of inflation with the
necessary emergence from the inflationary stage in the usual way: decay of scalar
fields, particle production, reheating etc. Moreover, we investigated the role of
the chiral metric component $h_{22}$ responsible for the interaction of the first phantom
chiral field and the second canonical field.

\section{Acknowledgments}

SC is thankful to the University of KwaZulu-Natal, the University of Zululand
and  the NRF for financial support and warm hospitality during his visit in 2011
to South Africa. SDM acknowledges that this
work is based upon research supported by the South African Research
Chair Initiative of the Department of Science and Technology and the
National Research Foundation.

\end{document}